**Microscale Hydrogen, Carbon, and Nitrogen Isotopic Diversity of Organic Matter in Asteroid Ryugu**

Larry R Nittler[a,b*], Jens Barosch[b,c], Katherine Burgess[d], Rhonda M Stroud[a,d], Jianhua Wang[b], Hikaru Yabuta[e], Yuma Enokido[f], Megumi Matsumoto[f], Tomoki Nakamura[f], Yoko Kebukawa[g,h], Shohei Yamashita[i], Yoshio Takahashi[i], Laure Bejach[j], Lydie Bonal[k], George D Cody[b], Emmanuel Dartois[l], Alexandre Dazzi[m], Bradley De Gregorio[d], Ariane Deniset-Besseau[m], Jean Duprat[n], Cécile Engrand[j], Minako Hashiguchi[o], A.L. David Kilcoyne[p†], Mutsumi Komatsu[q], Zita Martins[r], Jérémie Mathurin[j], Gilles Montagnac[s], Smail Mostefaoui[n], Taiga Okumura[t], Eric Quirico[k], Laurent Remusat[n], Scott Sandford[u], Miho Shigenaka[e], Hiroki Suga[v], Yasuo Takeichi[i], Yusuke Tamenori[v], Maximilien Verdier-Paoletti[n], Daisuke Wakabayashi[i], Masanao Abe[w], Kanami Kamide[e], Akiko Miyazaki[w], Aiko Nakato[w], Satoru Nakazawa[w], Masahiro Nishimura[w], Tatsuaki Okada[w], Takanao Saiki[w], Satoshi Tanaka[w], Fuyuto Terui[x], Tomohiro Usui[w], Toru Yada[w], Kasumi Yogata[w], Makoto Yoshikawa[w], Hisayoshi Yurimoto[y], Takaaki Noguchi[z], Ryuji Okazaki[o], Hiroshi Naraoka[o], Kanako Sakamoto[w], Shogo Tachibana[t], Sei-ichiro Watanabe[aa], Yuichi Tsuda[w]

*Corresponding Author: lnittler@asu.edu, School of Earth and Space Exploration, Arizona State University, Mail Code 6004, Tempe, AZ 85287, USA.

[a]School of Earth and Space Exploration, Arizona State University, Tempe, AZ 85287, USA.

[b]Earth and Planets Laboratory, Carnegie Institution for Science, Washington, DC 20015, USA.

[c]School of Geosciences, University of Edinburgh, Edinburgh, EH9 3FE UK

[d]Materials Science and Technology Division, U.S Naval Research Laboratory, Washington, DC 20375, USA.

[e]Department of Earth and Planetary Systems Science, Hiroshima University, Hiroshima 739-8526, Japan.

[f]Department of Earth Science, Tohoku University, Sendai 980-8578, Japan.

[g]Department of Earth and Planetary Sciences, Tokyo Institute of Technology, Tokyo 152-8551, Japan

[h]Division of Materials Science and Chemical Engineering, Yokohama National University, Yokohama 240-8501, Japan.

[i]Photon Factory, High Energy Accelerator Research Organization, Tsukuba 305-0801, Japan.

[j]Laboratoire de Physique des 2 Infinis Irène Joliot-Curie, Université Paris-Saclay, Centre National de la Recherche Scientifique, 91405 Orsay, France.

[k]Institute de Planétologie et d'Astrophysique, Université Grenoble Alpes, 38000 Grenoble, France.

[l]Institut des Sciences Moléculaires d'Orsay, Université Paris-Saclay, Centre National de la Recherche Scientifique, 91405 Orsay, France.

[m]Institut Chimie Physique, Université Paris-Saclay, Centre National de la Recherche Scientifique, 91405 Orsay, France.

[n]Institut de Mineralogie, Physique des Materiaux et Cosmochimie, Museum National d'Histoire Naturelle, Centre National de la Recherche Scientifique, Sorbonne Université, 75231 Paris, France.

[o]Department of Earth and Planetary Sciences, Kyushu University, Fukuoka 819-0395, Japan.

[p]Advanced Light Source, Lawrence Berkeley National Laboratory, Berkeley, CA 94720, USA.

[q]Department of Earth Sciences, Waseda University, Tokyo 169-8050, Japan.

[r]Centro de Química Estrutural, Institute of Molecular Sciences and Department of Chemical Engineering, Instituto Superior Técnico, Universidade de Lisboa, Lisboa 1049-001, Portugal

[s]École normale supérieure de Lyon, University Lyon 1, 69342 Lyon, France.

[t]Department of Earth and Planetary Science, University of Tokyo, Tokyo 113-0033, Japan.

[u]NASA Ames Research Laboratory, Moffett Field, CA 94035, USA.

[v]Japan Synchrotron Radiation Research Institute, Hyogo 679-5198, Japan.

[w]Institute of Space and Astronautical Science, Japan Aerospace Exploration Agency, Sagamihara 252-5210, Japan.

[x]Department of Mechanical Engineering, Kanagawa Institute of Technology, Atsugi 243-0292, Japan.

[y]Department of Earth and Planetary Sciences, Hokkaido University, Sapporo 060-0810, Japan.

[z]Department of Earth and Planetary Sciences, Kyoto University, Kyoto 606-8502, Japan.

[aa]Department of Earth and Environmental Sciences, Nagoya University, Nagoya 464-8601; Japan.

[†]Deceased

## Abstract

We report the H, C, and N isotopic compositions of microscale (0.2 to 1 μm) organic matter in samples of asteroid Ryugu and the Orgueil CI carbonaceous chondrite. Three regolith particles of asteroid Ryugu, returned by the Hayabusa2 spacecraft, and several fragments of Orgueil were analyzed by NanoSIMS isotopic imaging. The isotopic distributions of the Ryugu samples from two different collection spots are closely similar to each other and to the Orgueil samples, strengthening the proposed Ryugu-CI chondrite connection. Most individual sub-μm organic grains have isotopic compositions within error of bulk values, but 2-10% of them are outliers exhibiting large isotopic enrichments or depletions in D, $^{15}$N, and/or $^{13}$C. The H, C and N isotopic compositions of the outliers are not correlated with each other: while some organic grains are both D- and $^{15}$N-enriched, many are enriched or depleted in one or the other system. This most likely points to a diversity in isotopic fractionation pathways and thus diversity in the local formation environments for the individual outlier grains. The observation of a relatively small population of isotopic outlier grains can be explained either by escape from nebular and/or parent body homogenization of carbonaceous precursor material or addition of later isotopic outlier grains. The strong chemical similarity of isotopically typical and isotopically outlying grains, as reflected by synchrotron x-ray absorption spectra, suggests a genetic connection and thus favors the former, homogenization scenario. However, the fact that even the least altered meteorites show the same pattern of a small population of outliers on top of a larger population of homogenized grains indicates that some or most of the homogenization occurred prior to accretion of the macromolecular organic grains into asteroidal parent bodies.

## 1. Introduction

Many small primitive Solar System bodies, including comets and some asteroids, are rich in organic molecules and were likely important contributors of prebiotic materials to the early Earth (Martins et al., 2020). Their organic inventories range from insoluble macromolecular material to tens of thousands of identified soluble molecules, including prebiotic ones like amino acids (Alexander et al., 2017). The origin(s) of extraterrestrial organic matter are highly debated. Isotopic anomalies, mainly excesses in deuterium ($^{2}$H) and/or nitrogen-15 ($^{15}$N), point to an origin for at least some of the organic matter in the Sun's protosolar molecular cloud, but such primordial interstellar organic matter was likely mixed with material formed in the protoplanetary disk and modified during disk and planetary alteration processes during the early evolution of the Solar System. The characterization of organic matter from a known carbon-rich asteroid, with minimal terrestrial contamination, was a high priority of JAXA's Hayabusa2 mission, which returned 5.4 g of regolith from two different touchdown sites on asteroid 162173 Ryugu to Earth in December 2020 (Yada et al., 2022). While the first touchdown (Chamber A of the sample return canister) sampled the top surface, the second touchdown collection (Chamber C) is expected to have sampled sub-surface material excavated during the formation of an artificial crater by the Hayabusa2 Small Carry-on Impactor experiment (Arakawa et al., 2020).

Preliminary results from the Hayabusa2 Initial Analysis and Curation Teams have shown that Ryugu bears striking affinities to Ivuna-type (CI) carbonaceous chondrites, an extremely rare but scientifically important class of meteorites (Ito et al., 2022; Nakamura et al., 2022a; Nakamura et al., 2022b; Yokoyama et al., 2022). Organic matter makes up about 3 wt% of the samples and is similar in many respects to that seen in CI and other carbonaceous chondrites (Dartois et al., 2023; Naraoka et al., 2023; Yabuta et al., 2023). It is generally present as sub-micron particles and diffuse C associated with phyllosilicate minerals. Nanoscale analytical techniques have revealed an enormous diversity in functional group chemistry and isotopic compositions on small scales (e.g., Yabuta et al., 2023), indicating a rich and complex history. Yabuta et al. (2023) reported microscale H, C, and N isotopic data for a small number of Ryugu samples and insoluble organic matter (IOM) residues. These data were acquired with Cameca NanoSIMS instruments at the Museum National d'Histoire Naturelle (Paris) and the Carnegie Institution (Washington DC).

Here, we report an expanded study of the small-scale isotopic systematics of organic matter in additional intact Ryugu samples with the Carnegie instrument. The prior Carnegie data were acquired on thin slices of particles prepared by ultramicrotomy and placed on silicon wafers. In this work, we instead studied particle fragments that were pressed flat into clean annealed gold foils. Avoiding ultramicrotomy and using a cleaner substrate reduces the potential for terrestrial contamination affecting the isotopic measurements and thicker samples allow for more follow-up studies (e.g., by transmission electron microscopy, TEM) than possible with microtomed samples that have been analyzed by SIMS. We also studied fragments of the Orgueil CI chondrite to allow a direct comparison of Ryugu and CI samples prepared and analyzed by identical methods.

## 2. Material and methods

Two Ryugu particles were allocated to the Carnegie Institution for NanoSIMS analysis: A0108-13 from the first touchdown site and C0109-2 from the second. We used a fine needle attached to a micromanipulator to break small (20-30 μm) fragments off of these fluffy, fragile particles and transfer them to clean Au foils attached to Al stubs. We also transferred several similarly sized fragments from the Orgueil (CI1) carbonaceous chondrite. The Ryugu Chamber A, C, and Orgueil samples were each placed on a separate Au mount to avoid the possibility of cross-contamination. Grains were subsequently pressed flat into the Au foils with clean quartz disks (Fig. 1A). All samples were Au-coated and heated under vacuum to remove adsorbed water and nitrogen prior to NanoSIMS analysis. The pressed samples were imaged in a JEOL 6500F scanning electron microscope prior to NanoSIMS analysis (10 kV accelerating voltage, 1 nA beam current).

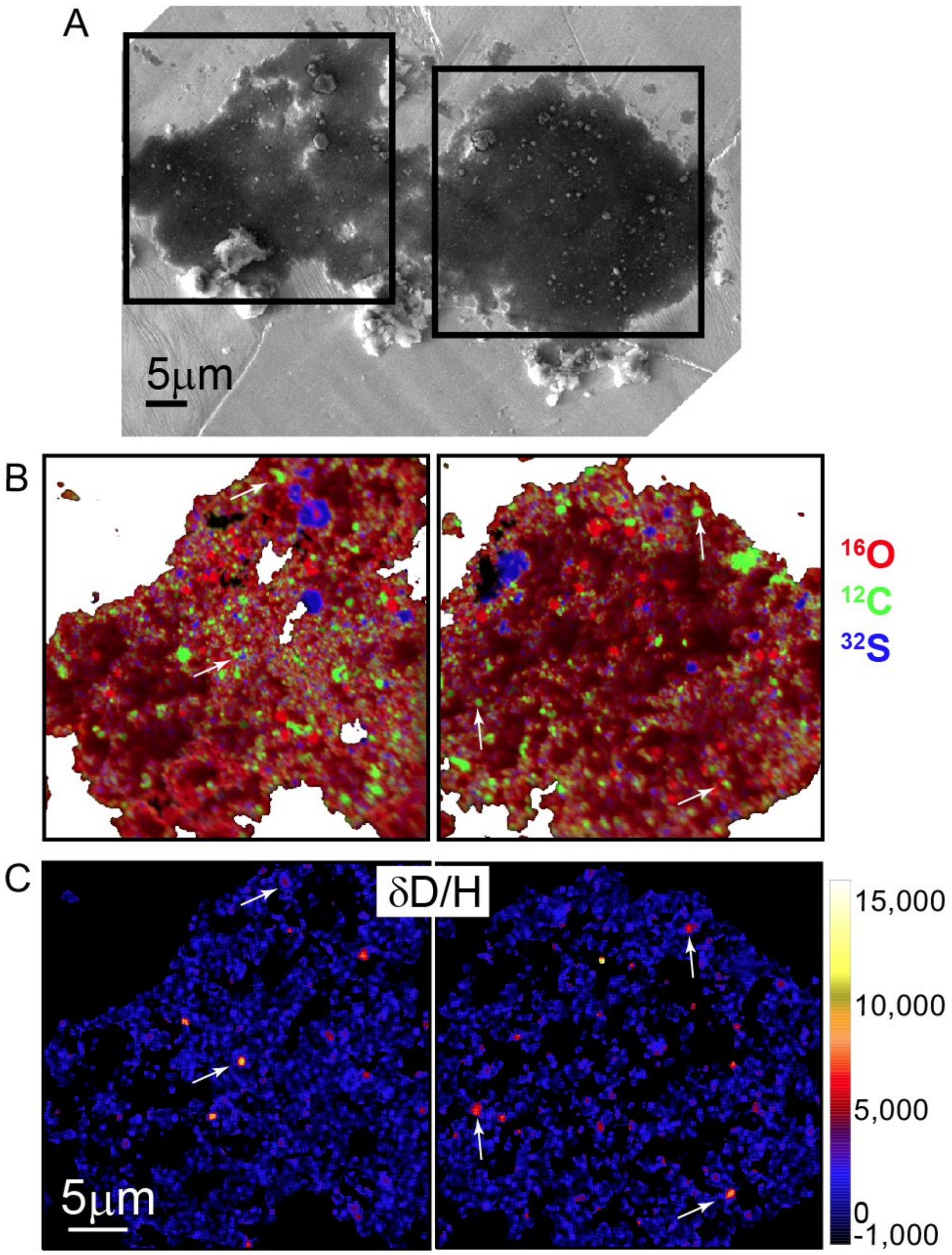

**Figure 1.** A) Secondary electron image of Ryugu fragment A0108-95 pressed into Au foil. Squares indicate locations of two NanoSIMS measurement locations. B) NanoSIMS composite RGB images of two imaged areas (Red= $^{16}O$, Green=$^{12}C$, Blue= $^{32}S$). Organic particles are visible as bright green regions. The arrows indicate organic particles highly enriched in deuterium. C) False-color NanoSIMS δD/H ratio maps of two imaged areas. Many organic grains (some indicated by arrows) have anomalously high ratios ("hotspots"). Delta-values are the part per thousand (‰) deviation of a ratio from a standard: $\delta R = 1000 \times (R_m/R_s - 1)$ where $R_m$ is a measured ratio and $R_s$ is the standard. Therefore, the color scale in panel C corresponds to D/H ratios ranging from zero to 16× the terrestrial standard value of $1.55 \times 10^{-4}$.

We analyzed the pressed fragments for their H, C, and N isotopic compositions with a Cameca NanoSIMS 50L ion microprobe at the Carnegie Institution in multicollector imaging mode (Fig. 1B, C). The NanoSIMS methods used were closely similar to those reported for the limited set of microtomed Ryugu samples by Yabuta et al. (2023). Each sample was measured in two sessions with a $Cs^+$ primary ion beam. In the first session ("C-N session"), we measured C and N isotopes (as $C_2^-$ and $CN^-$ molecules) along with $^{16}O^-$, $^{28}Si^-$, and $^{32}S^-$ ions, and in the second ("H session") we measured $^{1}H^-$, $^{2}H^-$, and $^{12}C^-$. The primary beam current was ~0.5-1 pA for the first and 1.5-3 pA for the second session, with corresponding spatial resolutions of ≈150 nm and ≈250 nm, respectively. The raster size, pixel resolution (e.g., 256×256 or 512×512 pixel), and the number of repeated measurement frames were varied according to the size of the sample to give similar spatial resolutions and similar total counting times per unit area. We used an organic solid with composition $C_{30}H_{50}O$ and IOM extracted from carbonaceous chondrite Queen Alexandra Range (QUE) 99177 to correct isotopic ratios for H and N instrumental mass fractionation and to determine C/N and C/H element ratios from secondary ion ratios. C isotopic ratios were found to depend on the count rate of $^{12}C_2^-$ ions, most likely reflecting so-called "quasi-simultaneous arrivals" (Slodzian et al., 2004). We thus empirically corrected for this effect (supplemental information) and corrected for instrumental fractionation under the assumption that the average composition of both the Ryugu and Orgueil sample organic matter is $\delta^{13}C = -17‰$, as seen for isolated IOM from CI chondrites (Alexander et al., 2007). This value is in good agreement with the measurements of Ryugu organic C by Nakamura et al. (2022a). Nonetheless, our procedure introduces some systematic uncertainty in the absolute C isotopic compositions. Upon analyzing the data, we discovered that the C-N images for Orgueil (which were acquired during a separate NanoSIMS analysis session) had lower secondary ion count rates than for the Ryugu samples, despite nominally similar analysis conditions, and hence larger statistical uncertainties. The cause of this difference is still undetermined.

The images were analyzed with the L'image software (L. Nittler, www.limagesoftware.net). For each Ryugu fragment, image frames were aligned to correct for beam or stage drift and corrected for the detector deadtime. For some images, background electron multiplier noise was visible in the D ion images. This background (which ranged from 0.008 to 0.015 counts per second) was determined by summing background image pixels outside the samples themselves and the derived D/H ratios were corrected accordingly. We defined two sets of individual C-rich regions of interest (ROIs) in the images. First, we used an automated particle definition algorithm (Nittler et al., 1997) to define as many ROIs as possible in the $^{12}C_2$ images acquired during C-N measurements, down to the smallest size visible in the images (~150 nm) in order to estimate the size distribution of C-rich particles above this limit. Second, we used a combination of automated and manual techniques to define those grains large enough (generally >200 nm) with high enough signals to allow isotopic measurement in the $C_2$ and C images from both the C-N and H isotope sessions, respectively. Pixels were assigned to an ROI if the C signal ($C_2^-$ or $C^-$) was greater than 50% of the peak intensity for that ROI. Comparison of these images was then used to manually correlate individual ROIs across the H and C-N measurements where possible (Fig. S1). Because SIMS is a destructive technique, however, some ROIs found in the initial C and N session were gone by the time of the H measurements, while some new C-rich grains were exposed in the latter.

In addition to the carbonaceous ROIs, we defined "bulk" ROIs for each image. For C and N these were defined by summing all the pixels with $^{16}O^-$ or $^{12}C_2^-$ secondary ion signals greater than 5% of their maximum intensity. For H, we used the same relative threshold for $H^-$ counts.

We also analyzed a thin section of Ryugu particle C0002 prepared by focused ion beam (FIB) at the Tohoku University and first analyzed by C-edge X-ray Absorption Near-Edge Spectroscopy (XANES) at the Photon Factory, High Energy Accelerator Research Organization in Japan. Details of the sample preparation and STXM measurements can be found in Yabuta et al. (2023). For NanoSIMS analysis, we used a Thermo Fisher Helios G3 DualBeam FIB at the US Naval Research Lab to remove the section from the grid where it had originally been attached, transfer it to a Au foil, and weld its edges to the Au with Pt. NanoSIMS analysis of C and N isotopes in the sections followed the same procedures as used for the pressed fragments. However, we found that FIB-deposited C contamination surrounding the section showed strongly varying instrumental fractionation for C isotopes. We thus consider the C data suspect and only report N isotopic data for this sample.

## 3. Results

### *3.1 Carbonaceous grain size and morphology*

We analyzed a total of ~3700 μm$^2$, ~3800 μm$^2$, and ~7600 μm$^2$ of our pressed A0108-13, C0109-2, and Orgueil particles, respectively (Table 1). Previous high spatial-resolution investigations have shown organic matter to be present in Ryugu both as particles ranging in size from 10s of nanometers to microns (Ito et al., 2022; Yabuta et al., 2023) and as diffuse C associated with phyllosilicate grains. The measured size distribution of >11,000 C-rich grains in our Ryugu pressed-grain samples (black histogram in Fig. 2) shows a peak between 200 and 300 nm, but with rare particles ranging up to 2 microns. The sharp lower cutoff at ~150 nm reflects the sharply decreasing detection efficiency for grains below the spatial resolution of the images; many additional smaller carbonaceous grains are known to be present as well, based on TEM and high-resolution infrared analysis (Yabuta et al., 2023; Stroud et al., 2024).

For isotopic analysis, we defined a much more limited set of C-rich ROIs in the images: 803 in A0108-13, 418 in C0109-2, and 1012 in Orgueil (see Supplementary Information for a discussion of how ROIs were selected). A small number of the C-rich ROIs were found to be extremely isotopically anomalous in C isotopes ($^{13}$C-enriched or depleted) indicating an origin as presolar grains of SiC or graphite that condensed in the outflows of ancient stars. These have already been reported by Barosch et al. (2022) and we do not repeat their discussion here. The remainder of the particles are inferred to be organic in nature, based on the comparison with other nanoscale characterizations of related Ryugu samples (Yabuta et al., 2023) which have shown the ubiquity of organic matter in the samples. As seen in Fig. 2, organic grains in the two Ryugu touchdown site samples and the Orgueil samples show very similar size distributions, though the Orgueil images show a slightly smaller average ROI size (365±4 nm compared to 416±5 nm, 416±7 nm for the A and C data, respectively; errors are one standard error of the mean). These distributions are shifted to significantly larger sizes than the 200- to 250-nm typical size of C-rich grains in the images, reflecting the bias towards larger grains to obtain signals statistically sufficient to determine isotopic ratios. The total area of the ROIs with isotopic data is about a quarter of the total area of defined ROIs, and thus represent at most about 25% of the total organic C present in the Ryugu and Orgueil samples; the remainder is in smaller grains or diffuse C.

**Table 1:** Bulk isotopic compositions of pressed Ryugu and Orgueil particles

| Image Name | Sample[a] | Area ($\mu m^2$) | δD/H[b] (‰) | δD/H[c] (‰) | EM bkg[d] (cps) | $\delta^{15}N/^{14}N$[b] (‰) | $\delta^{13}C/^{12}C$[b] (‰) |
|---|---|---|---|---|---|---|---|
| HY2_A4-02a | A0108-92 | 529 | 397±49 | 480±55 | | -26±8 | -10.9±12.0 |
| HY2_A4-02b | A0108-92 | 207 | 319±50 | 282±54 | 0.011 | 26±11 | -4.2±12.1 |
| HY2_A4-02c | A0108-92 | 443 | 497±56 | 576±64 | 0.011 | 14±8 | -16.5±12.0 |
| HY2_A4-03a | A0108-93 | 287 | 290±47 | 265±49 | 0.011 | 45±8 | -20.4±12.0 |
| HY2_A4-06a | A0108-94 | 286 | 452±49 | 487±52 | | 5±6 | -24.4±11.9 |
| HY2_A4-06b | A0108-94 | 316 | 558±62 | 723±73 | | 34±6 | -26.8±11.9 |
| HY2_A4-08a | A0108-95 | 839 | 565±51 | 632±53 | 0.015 | 24±5 | -12.0±12.0 |
| HY2_A4-08b | A0108-95 | 805 | 519±49 | 647±54 | 0.003 | 22±3 | -17.4±12.0 |
| Average | A0108-13 | | 450±105 | 511±168 | | 18±21 | -16.6±7.4 |
| HY1_A2-03 | C0109-84 | 423 | 706±59 | 1172±79 | | 114±15 | -36.4±12.3 |
| HY1_A3-01 | C0109-85 | 234 | 454±56 | 553±62 | 0.005 | 44±16 | -9.1±12.3 |
| HY1_A3-02 | C0109-86 | 491 | 1081±73 | 1367±84 | | 47±9 | -20.2±12.1 |
| HY1_A3-03 | C0109-87 | 199 | 361±49 | 434±54 | | 41±11 | -22.9±12.2 |
| HY1_A3-04 | C0109-88 | 77 | 717±67 | 820±74 | 0.005 | 41±15 | -12.3±12.0 |
| HY1_A3-05 | C0109-89 | 168 | 448±56 | 532±62 | | -29±16 | -12.1±12.2 |
| HY1_A3-07 | C0109-90 | 183 | 318±56 | 521±74 | | 77±19 | -13.3±12.5 |
| HY1_A3-09 | C0109-91 | 503 | 225±57 | 331±71 | | 28±13 | -7.9±12.2 |
| HY1_A3-10 | C0109-92 | 609 | 576±59 | 819±73 | 0.008 | 39±10 | 3.8±12.1 |
| HY1_A3-11 | C0109-93 | 327 | 419±58 | 928±91 | | 84±13 | -6.8±12.2 |
| HY1_A3-17 | C0109-94 | 429 | 111±48 | 324±65 | | 37±11 | -6.7±12.1 |
| HY1_A3-20 | C0109-95 | 178 | 512±64 | 635±75 | | 36±16 | -26.3±12.1 |
| Average | C0109-2 | | 433±256 | 627±328 | | 47±35 | -12.1±10.7 |
| OrgM_A3-01 | Orgueil | 830 | 352±66 | 907±99 | | 39±9 | -20.7±13.4 |
| OrgM_A3-02 | Orgueil | 385 | 330±64 | 480±75 | | 24±7 | -33.5±13.2 |
| OrgM_A3-03 | Orgueil | 730 | 356±66 | 621±83 | | 23±7 | 2.4±13.5 |
| OrgM_A3-05 | Orgueil | 312 | 389±67 | 541±76 | | 17±6 | -21.7±13.4 |
| OrgM_A3-06 | Orgueil | 482 | 309±63 | 798±89 | | 85±8 | -19.7±13.5 |
| OrgM_A3-07 | Orgueil | 298 | 360±66 | 531±76 | | 34±7 | -16.4±13.5 |
| OrgM_A4-01 | Orgueil | 1278 | 457±70 | 882±93 | | 41±6 | -39.6±13.4 |
| OrgM_A4-02 | Orgueil | 558 | 304±64 | 364±71 | | 1±9 | -10.1±13.7 |
| OrgM_A4-03 | Orgueil | 603 | 327±64 | 481±74 | | 7±9 | -14.1±13.6 |
| OrgM_A4-04 | Orgueil | 677 | 452±70 | 680±83 | | 16±6 | -22.9±13.4 |
| OrgM_A4-05 | Orgueil | 380 | 397±68 | 681±84 | | 30±8 | -19.4±13.5 |
| OrgM_A5-01 | Orgueil | 250 | 308±65 | 319±68 | | 25±13 | -17.7±13.8 |
| OrgM_A5-02a | Orgueil | 341 | 335±65 | 580±83 | | 24±13 | 2.7±13.8 |
| OrgM_A5-02b | Orgueil | 465 | 441±70 | 929±99 | | 35±7 | -6.9±13.7 |
| Average | Orgueil | 7588 | 365±54 | 628±195 | | 29±20 | -17.0±11.7 |

[a]Pressed fragments of Ryugu A0108-13 and C0209-2 were assigned new sample numbers as shown
[b]Value for bulk ROIs based on $^{16}O$ or H ion signal greater than 5% of maximum intensity.
[c]Value for bulk ROIs based on C ion signal greater than 5% of maximum intensity.
[d]Electron Multiplier background (D images corrected).

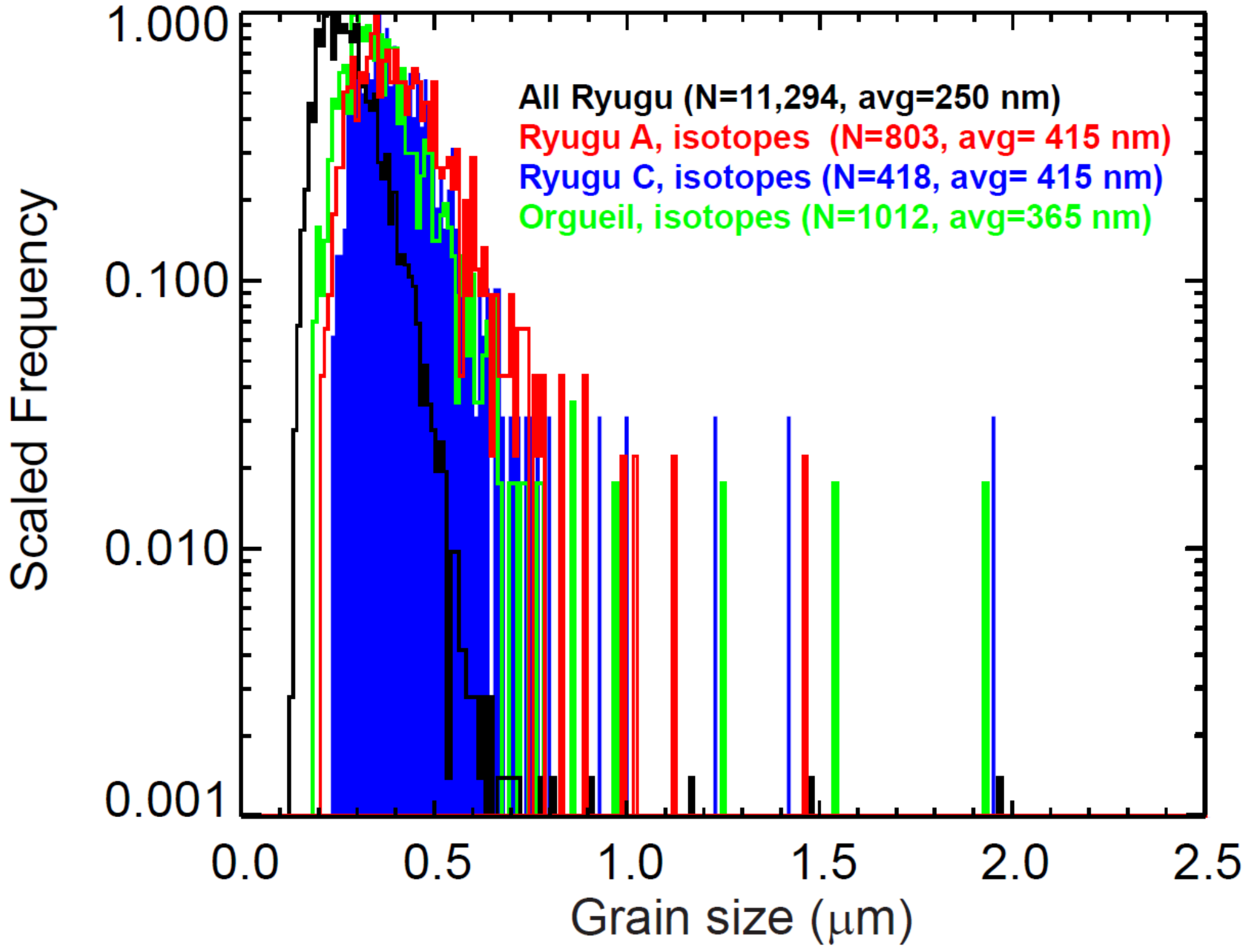

**Figure 2.** Histograms of the diameters of C-rich regions of interest (ROIs) defined in the NanoSIMS images. The black histogram represents all ROIs visible in the Ryugu images. The colored histograms represent filtered sets of ROIs from the Ryugu and Orgueil samples used for isotopic analysis; counting statistics limit these to grains >200 nm.

The large majority of organic grains visible in the Ryugu and Orgueil NanoSIMS images appear round in shape (e.g., Fig. 1) and are likely in many cases to be nanoglobules, sometimes-hollow carbonaceous spheroids previously reported in carbonaceous chondrites and comet Wild-2 samples (De Gregorio et al., 2010; De Gregorio et al., 2013; Nakamura-Messenger et al., 2006). TEM analysis of Ryugu samples has also revealed nanoglobules to be the dominant form of particulate organic matter (Stroud et al., 2024). However, occasionally grains with unusual morphologies are observed (e.g., Fig. 3) including nanoglobule clusters ("bunches-of-grapes"), vermiform, and other shapes. Indicated by the dotted line in Fig. 3D is a highly unusual blocky or needle-like C-rich region with sharp rectangular edges. Although the overall C, H, and N signals were low, this grain has among the highest enrichments of D and $^{15}N$ seen in our samples so far (Figs. 3E,F; Section 3.3). We used the NRL FEI Helios G3 $Ga^+$ FIB-SEM to prepare an ultrathin section of this Ryugu fragment (C0109-A03-02), crossing the blocky "grain." TEM analysis of the section did not reveal obvious carbonaceous material in the location of the grain. Although this grain may have been initially thicker and completely consumed by the NanoSIMS prior to TEM analyses, we suspect that it was more likely a very thin isotopically anomalous C coating on minerals and its shape reflects that of a pre-existing cavity between grains. A thin coating is consistent with the overall low organic ion signals in this ROI.

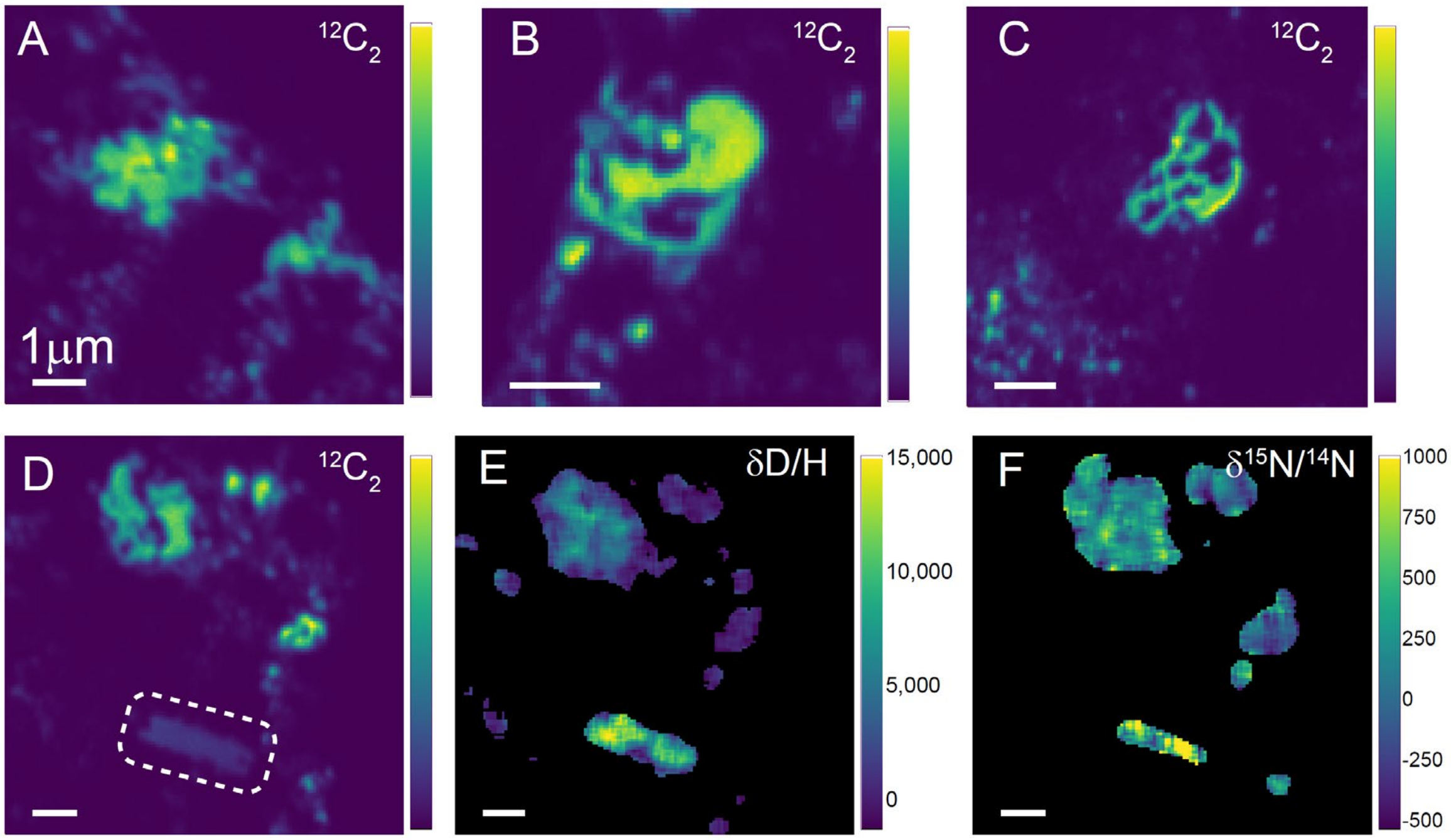

**Figure 3:** Diversity of morphologies for organic particles in Ryugu and Orgueil. A) NanoSIMS $^{12}C_2$ ion image of a portion of fragment A0108-95 (Fig. 1). B) NanoSIMS $^{12}C_2$ ion image of a portion of fragment A0108-92. C) NanoSIMS $^{12}C_2$ ion image of a portion of Orgueil fragment Org-A03-06. D) NanoSIMS $^{12}C_2$ ion image of a portion of fragment C0109-86; the dashed box indicates a faint, block-shaped region. E) NanoSIMS δD/H and F) $\delta^{15}N/^{14}N$ maps of region from D. The blocky grain is highly enriched in both D and $^{15}N$. Scale bars are one micron.

### *3.2 Bulk Hydrogen, Carbon, and Nitrogen Isotopes*

The H and N isotopic compositions for the "bulk" ROIs of our pressed samples are provided in Table 1 and shown in Fig. 4. Note that even with the ultrahigh vacuum (few $10^{-10}$ torr) and the pre-heating of samples, some surface contamination by terrestrial hydrogen is inevitable under the low-current, imaging mode we used (Stephant et al., 2014), so all reported non-terrestrial D/H ratios should be considered lower limits. An additional potential source of systematic uncertainty in these data arises from the fact that we did not analyze any inorganic H-isotope standards, yet the bulk ROIs include both organic matter and silicates. However, matrix effects are generally similar for H isotopes from organic and inorganic phases under $Cs^+$ bombardment (Aléon et al., 2001; Piani et al., 2018).

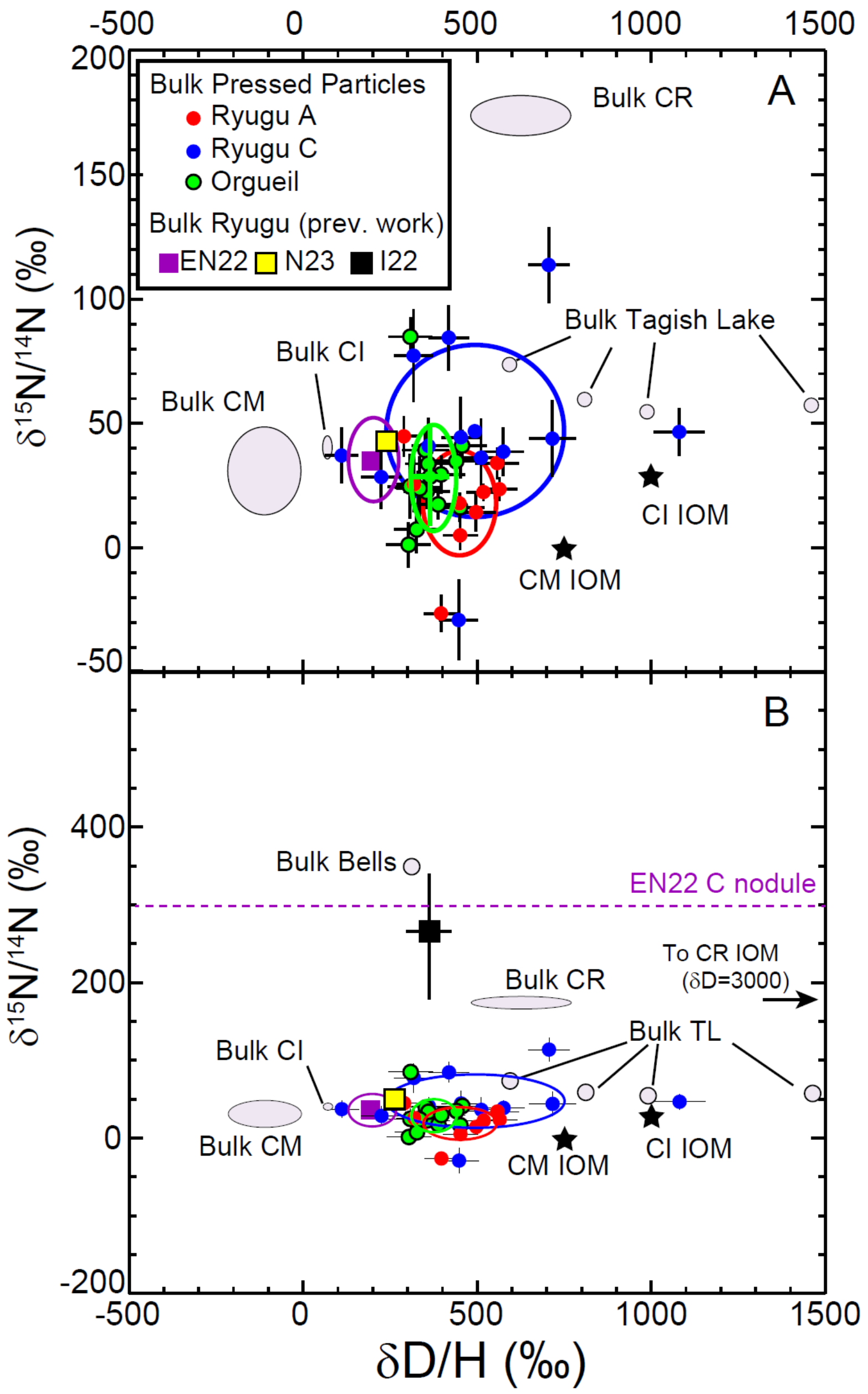

**Figure 4.** Bulk H and N isotopes for Ryugu and Orgueil samples. The filled circles represent individual NanoSIMS measurements, while corresponding colored ellipses indicate average and one standard deviation of the measurements. The Ryugu Chamber A and C and Orgueil samples have overlapping compositions. Filled ellipses and stars represent the compositions of bulk meteorites (Alexander et al., 2012; Herd et al., 2011) and insoluble organic matter (IOM) residues (Alexander et al., 2007). Filled squares indicate bulk Ryugu measurements by Nakamura et al. (2022a; EN22), Naraoka et al. (2023; N23), and Ito et al. (2022; I22). The dashed line in panel B indicates the $\delta^{15}N$ value for a "Carbonaceous nodule" reported by Nakamura et al. (2022a).

There is considerable scatter in bulk isotopes between the pressed fragments, reflecting heterogeneous distributions of small organic particles with extreme isotope anomalies (Sec. 3.3). Nevertheless, the Ryugu Chamber A, Ryugu Chamber C, and Orgueil samples all show average bulk compositions that agree with each other within one standard deviation (error ellipses), with slightly enriched D and $^{15}$N, relative to terrestrial values. The average D/H ratios lie between those of bulk CI chondrites (Alexander et al., 2012) and IOM derived from the same meteorites (Alexander et al., 2007). This does not indicate that our Ryugu and Orgueil samples are D-rich compared to other CI chondrites, but rather the fact that our NanoSIMS conditions favor detection of H from organic matter over that from the phyllosilicates that dominate the samples' mineralogy (Deloule and Robert, 1995). In fact, limiting the "bulk" ROI definition to C-rich pixels to exclude phyllosilicates only slightly shifts the measured D/H ratios upward (Table 1). The bulk C isotopic ratios (Table 1) show much smaller variations among the different fragments than do the H or N isotopes. This directly reflects the much smaller range of C isotopic ratios among sub-μm organic grains compared to H and N isotopic ratios (Sec. 3.3).

### *3.3 Microscale Isotope Systematics*

The H, N, and C isotopic compositions and atomic N/C ratios of individual C-rich regions of interest (ROIs) from the Ryugu and Orgueil samples are plotted as histograms in Fig. 5. There are no clear differences in isotopic composition between the two Ryugu samples and the Orgueil grains. As seen previously for Ryugu samples (Yabuta et al., 2023), for all three isotopic ratios, the vast majority of the ROI data are distributed approximately normally around the bulk compositions, with a small fraction of ROIs being outliers (commonly referred to as "hotspots" or "coldspots," depending on whether the rare isotope is enriched or depleted). Whether a specific grain can be identified as an outlier depends on its measurement precision, which is highly variable among the ROIs due to a wide range of secondary ion count rates. We define an outlier here as an ROI whose isotopic ratio differs from the bulk value for the dataset by $>3\sigma$, where $\sigma$ is the uncertainty for the ROI. With this definition, isotopic outliers make up less than 10% of the ROIs (by number) for Ryugu A, C, and Orgueil samples (Table 2). The absolute D/H and $^{15}$N/$^{14}$N isotopic ratios for outliers span wide ranges, factors of ~50 and 3, respectively. Altogether, the small fractions of outliers emphasize that despite the obvious presence of extreme isotopic anomalies (e.g., Fig. 1C) in all the images, a primary result of these measurements is that sub-μm organic particles in Ryugu show a mostly homogeneous isotopic composition.

The distributions of N/C ratios among the C-rich grains are similar but not identical between the Ryugu and Orgueil samples (Fig. 5D). The three histograms peak between N/C=0.015 and 0.02, with the Orgueil ROIs shifted on average to slightly higher values. These values are considerably lower than the value measured for purified IOM from CI chondrites (0.035; Alexander et al., 2007). This may indicate a difference between the bulk organic matter in the samples compared to IOM residues and that C-rich particles may not be constituted of pure IOM but rather a mixture of insoluble and soluble organic matter. However, it may also indicate an error with inadequate matrix-matching since the standard used to convert $CN/C_2$ secondary ion ratios to N/C ratios was a purified organic sample compared to the mixed silicate-organics nature of our samples. Although most ROIs show relatively low N/C comparable to carbonaceous chondrite IOM, some rare (<4%) ROIs reach much higher values (up to 0.08) indicating that they are unusually N-rich.

**Table 2.** Statistics for isotopic outliers in Ryugu and Orgueil samples.

| Sample | δD/H (‰) | | | | $\delta^{15}$N (‰) | | | | $\delta^{13}$C (‰) | | | |
|---|---|---|---|---|---|---|---|---|---|---|---|---|
| | ROI% | area% | min | max | # % | area% | min | max | ROI% | area% | min | max[a] |
| Ryugu A0108-13 | 9.6 | 0.43 | -724 | 13,735 | 6.8 | 0.44 | -780 | 1090 | 1.9 | 0.09 | -280 | 140 |
| Ryugu C0109-2 | 8.2 | 0.32 | -600 | 14,890 | 6 | 0.26 | -670 | 1210 | 2.7 | 0.027 | -340 | 110 |
| Ryugu C0002 FIB | | | | | | 11[b] | | | | | | |
| Orgueil | 9.5 | 0.27 | -590 | 12,230 | 3.6 | 0.15 | -405 | 1070 | 1.0 | 0.022 | -390 | 190 |
| Ryugu A0040 matrix[c] | | | | | | 0.3 | | | | 0.02 | | |
| Ryugu A0040 clast 2[c] | | | | | | 1.1 | | | | 0.2 | | |
| Ryugu C0002 matrix[c] | | | | | | 0.07 | | | | 0.02 | | |
| Ryugu C0002 clast 1[c] | | | | | | 1.4 | | | | 0.3 | | |

[a]Excludes highly $^{13}$C-enriched presolar grains reported by Barosch et al. (2022)
[b]Ratio of area of $^{15}$N-rich grains to uncontaminated part of section (Fig. 8)
[c]From Nguyen et al. (2023)

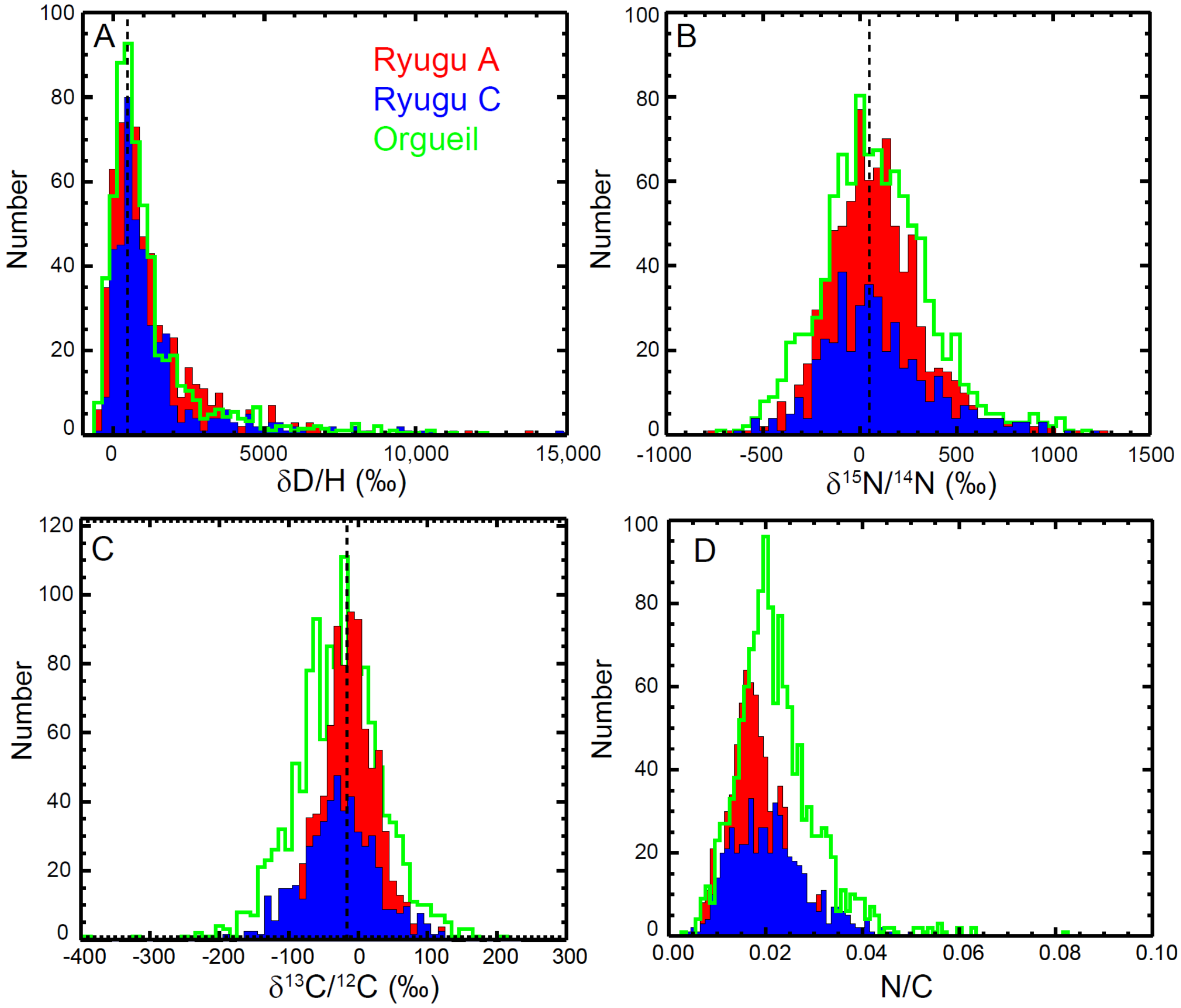

**Figure 5.** Histograms of isotopic and elemental ratios derived for C-rich ROIs in Ryugu and Orgueil particles. The vertical dashed lines indicate average values for the combined data sets. A) δD/H values. B) $\delta^{15}N/^{14}N$ values; one Orgueil ROI with $\delta^{15}N$=2100±600 ‰ is excluded. C) $\delta^{13}C/^{12}C$ values. D) N/C atomic ratios.

In Figure 6, we plot the δD/H versus $\delta^{15}N/^{14}N$ values for the ROIs that could be jointly identified in both the C-N and H-D imaging runs; only ROIs that are outliers in one or both isotopic ratios are shown. Recall that some 90% of the analyzed C-rich grains lie close to the bulk values, indicated by dashed lines. The ranges of D/H and $^{15}N/^{14}N$ ratios for the outliers is essentially identical to that seen in the more limited NanoSIMS data reported by Yabuta et al. (2023). As in the previous study, the H and N isotope ratios are uncorrelated. Although some outliers are enriched in both D and $^{15}N$, many are enriched or depleted in one of the isotopes, but not the other. Also, there is again no significant difference between either of the Ryugu touchdown samples and the Orgueil samples.

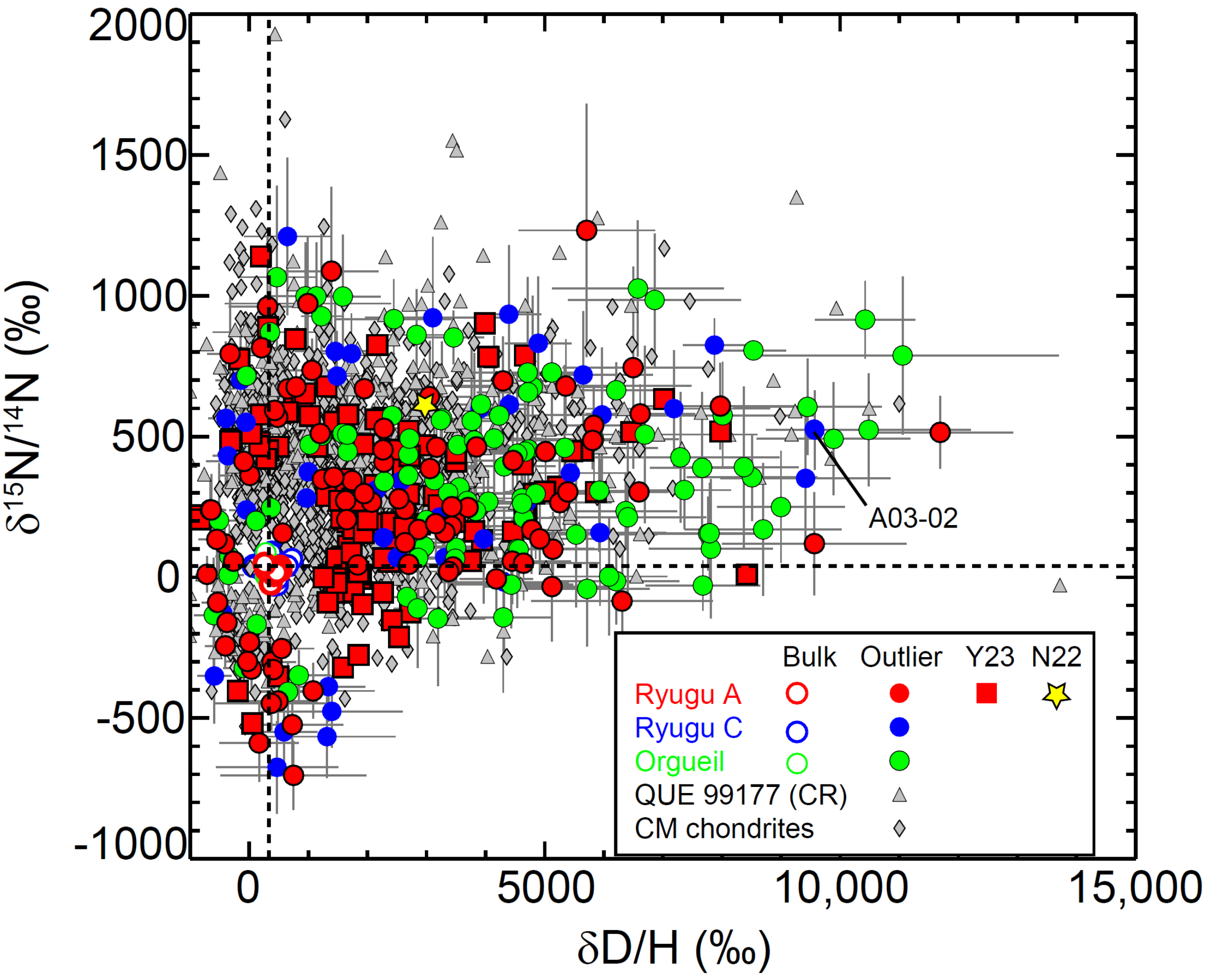

**Figure 6.** H and N isotopic ratios of isotopically anomalous organic ROIs in Ryugu and Orgueil compared to carbonaceous chondrites. The dashed lines indicate the average bulk values of the samples. Some 90% of all ROIs lie within errors of these values. The "Y23" data are those reported for microtomed chamber A particles by Yabuta et al. (2023). The QUE 99177 CR chondrite data were acquired in part by Nguyen et al. (2010). The CM chondrite data are from the least altered CM chondrites Paris, Asuka 12236, and Asuka 12169 (Nittler et al., 2021).

As seen in Figure 5C, the C isotopic ratio spans a narrower range for the Ryugu and Orgueil particles than do H or N isotopes. Nevertheless, grains with both modest and extreme $^{13}C$ enrichments and depletions are present. Excluding the most extreme anomalies associated with presolar stardust grains (Barosch et al., 2022), we plot the $\delta^{13}C/^{12}C$ values of outlier ROIs versus N and H isotopic ratios in Fig. 7. Again, there is no substantial difference between the Ryugu and Orgueil samples. Moreover, there is no correlation between C isotopes and either H or N isotopes; both $^{13}C$-enriched and $^{13}C$-depleted grains show a wide range of D/H and $^{15}N/^{14}N$ ratios, both enriched and depleted relative to the bulk. There is a moderate excess of $^{13}C$-depleted grains (72%) compared to $^{13}C$-enriched grains (28%) in our samples, but the difference is not very significant.

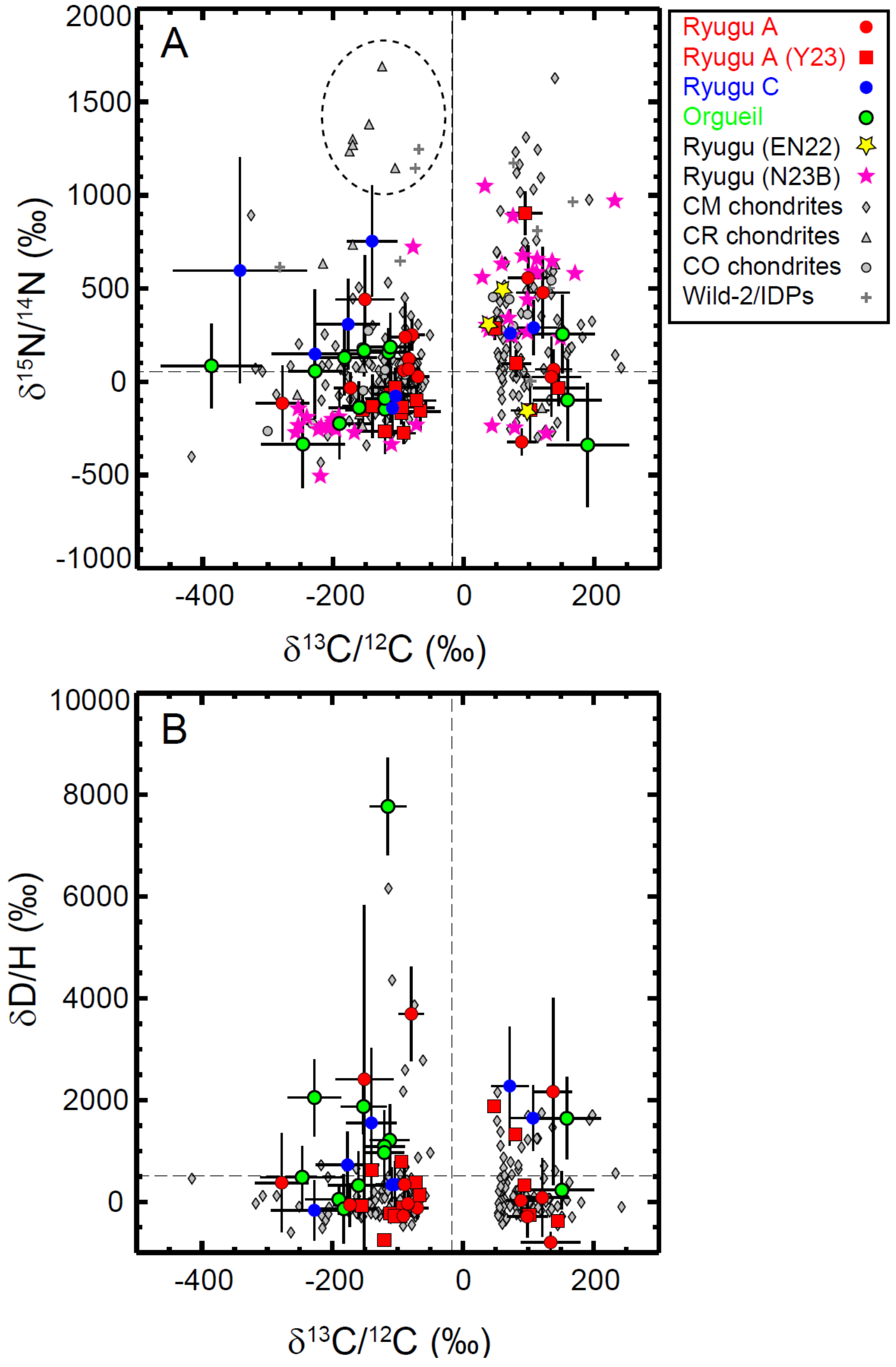

**Figure 7.** H, C, and N isotopic ratios for C-anomalous outlier ROIs in Ryugu and Orgueil samples compared to other primitive astromaterials. The dashed lines indicate average values of our Ryugu dataset. A) $\delta^{15}N/^{14}N$ values plotted versus $\delta^{13}C/^{12}C$ values. B) $\delta D/H$ values plotted versus $\delta^{13}C/^{12}C$ values. The "Y23" data are those reported for microtomed chamber A particles by Yabuta et al. (2023). EN22 and N23B data are Ryugu measurements from Nakamura et al. (2022a) and Nguyen et al. (2023), respectively. CM chondrite data include Bells (De Gregorio et al., 20013), Maribo (Vollmer et al., 2020), Paris, Asuka 12236, and Asuka 12169 (Nittler et al., 2021). CO data include Allan Hills 77307 (Bose et al., 2012) and Dominion Range 08006 (Nittler et al., 2018). CR data include QUE 99177 and Meteorite Hills 00426 (Floss and Stadermann, 2009) and GRO 95577 (De Gregorio et al., 2013). "Wild-2/IDPs" include a nanoglobule from comet Wild-2 (De Gregorio et al., 2010) and organic particles in stratospheric interplanetary dust particles (Floss et al., 2004; Floss et al. 2006; Busemann et al., 2009). The dotted ellipse indicates a population of $^{13}C$-poor, $^{15}N$-rich grains seen in other samples, but not yet identified in Ryugu.

Results from the correlated STXM-NanoSIMS study of the FIB section from Ryugu particle C0002 are shown in Figure 8. Although C contamination added during the mounting of the FIB section onto the SIMS mount covered part of the section (Fig. 8A), we were able to obtain N isotope data for 64 individual C-rich ROIs in the uncontaminated portion. Six of these correspond to grains for which C-XANES spectra were obtained (Fig. 8C). Error bars are large for most of the ROIs but strikingly, all six ROIs with XANES data are relatively enriched in $^{15}N$, with $\delta^{15}N$ ranging from 175±67‰ to 370±85‰. These grains make up 11% of the uncontaminated area of the section (Table 2). The C-XANES spectra (Fig. 8C) are similar to each other, showing absorptions due to aromatic carbon (285.2 eV), ketone functional groups (286.7 eV), and carboxyl functional groups (288.7 eV). They are similar to typical IOM spectra from CI chondrites, though they show a somewhat higher ratio of aromatic to ketone absorption. None appear similar to the "Highly Aromatic" spectra defined for some unusual organic grains in Ryugu (Yabuta et al., 2023). An additional C-XANES spectrum of a region of diffuse carbon in this section showed a molecular carbonate feature as seen in other Ryugu samples (Yabuta et al., 2023). Unfortunately, this region was covered by contamination during the process of transferring the sample for NanoSIMS analysis and thus no isotopic data are available for it.

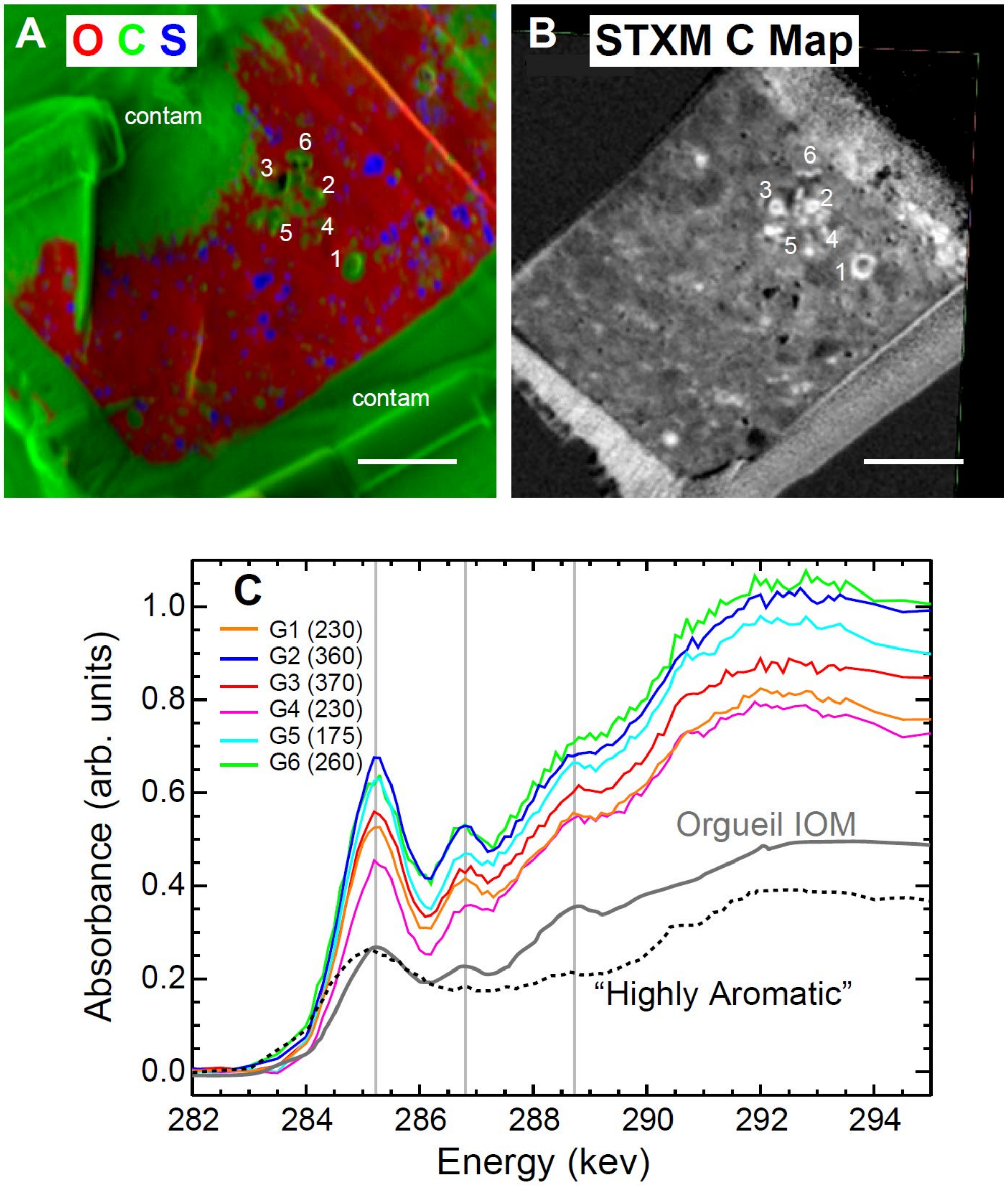

**Figure 8.** Correlated analysis of FIB section from Ryugu particle C0002. A) NanoSIMS composite RGB image (Red= $^{16}O$, Green=$^{12}C$, Blue= $^{32}S$). The section is surrounded and partially covered by C-rich contamination from the mounting process. Organic particles within the section are clearly visible as bright green regions. B) STXM map of C in section produced by ratioing absorption at 292 eV by that at 280 eV. Scale bars in A and B are 2 μm. C) C-XANES absorption spectra (-ln($I/I_0$) where I is the transmitted x-ray intensity and $I_0$ is the incident intensity) of the six C-rich grains labeled in panels A and B. The grey vertical lines correspond to photoabsorptions by aromatic carbon (285.2 eV), ketone functional groups (286.7 eV), and carboxyl functional groups (288.7 eV). Nitrogen isotopic compositions ($\delta^{15}N$ values) are given in parenthesis. Shown for comparison are spectra for insoluble organic matter (IOM) extracted from the Orgueil CI chondrite and a "Highly aromatic" spectrum measured in a different Ryugu sample (A0108-11a from Yabuta et al., 2023).

## 4. Discussion

### *4.1. Comparison with carbonaceous chondrites and other Ryugu studies*

Our results confirm the previous work presented in Yabuta et al. (2023) showing that Ryugu samples are in bulk moderately D- and $^{15}$N-rich, relative to terrestrial ratios, and that most sub-μm carbonaceous particles (>200 nm) are similar to the bulk, with up to 10% of them exhibiting anomalous isotopic enrichments or depletions (hotspots and coldspots). By analyzing Orgueil samples under essentially identical conditions, our data also allow a direct Ryugu-CI chondrite comparison. That the microscale H, C, and N isotopic distributions are essentially identical across these samples provides further evidence for the already-strong link between Ryugu and CI chondrites (Ito et al., 2022; Nakamura et al., 2022a; Nakamura et al., 2022b; Yokoyama et al., 2022).

Our data are also largely consistent with other Ryugu studies. For example, Nakamura et al. (2022a) studied Ryugu samples both in bulk and in situ by large-geometry SIMS. Their bulk N isotopic composition ($\delta^{15}$N=35‰; "N22" in Fig. 4) is in excellent agreement with our measured bulk values ($\delta^{15}$N≈35‰, Table 1) for Ryugu and Orgueil, especially considering that our data represent a much smaller amount of material. The Nakamura et al. (2022a) bulk D/H ratio is at the low end of our values, but this can be explained by the preference of SIMS to detect organic H, as discussed earlier (cf. Section 3.2). Naraoka et al. (2023) also found closely similar values to ours, $\delta$D/H=250‰ and $\delta^{15}$N=45‰ for a 38.4 mg aggregate Chamber A sample ("N23" in Fig. 4). However, Okazaki et al. (2023) found lower bulk $\delta^{15}$N values of 0 to 20‰, suggesting that there is isotopic heterogeneity at the mg scale among Ryugu samples. We see a notable exception to the agreement across Ryugu studies when considering the study of Ito et al. (2022). These authors used NanoSIMS to measure the H and N isotopes in seven FIB sections from three Ryugu particles. While the measured bulk $\delta$D/H values for these sections (200 to 600‰) are consistent with our values, the $\delta^{15}$N values are considerably higher on average ("I22" in Fig. 4B), with 4 out of 7 showing $\delta^{15}$N >150‰, up to 550‰, and an average value of 260‰. The sizes of the FIB sections analyzed by Ito et al. (2022) are not specified, but the one shown in their paper is comparable in size to our individual NanoSIMS images of pressed fragments, none of which shows similarly high bulk $^{15}$N enrichments. However, in their SIMS analysis of large areas of Ryugu samples, Nakamura et al. (2022a) identified a ≈20-μm C-enriched clast ("carbonaceous nodule") with $\delta^{15}$N=312 ‰ (Fig. 4B), as well as other ~10-μm concentrations of isotopically anomalous organic matter (e.g., yellow star in Fig. 6), showing that such materials exist, albeit rarely. Therefore, the unusually high bulk $^{15}$N contents reported for FIB sections by Ito et al. (2022) may reflect fortuitous sampling.

Aside from the previous work by Yabuta et al. (2023), there are as yet no comparable correlated H, C, and N isotope studies of Ryugu samples at the sub-μm scale and few such studies of meteorites with which to directly compare our results. Previous uncorrelated studies have revealed similar ranges of outlier compositions for H, C, and/or N isotopes to those seen here for CI, CR, CM, and CO chondrites (Bose et al., 2012; Busemann et al., 2006; Floss and Stadermann, 2009; Nittler et al., 2018; Nittler et al., 2019; Remusat et al., 2019; Remusat et al., 2010; Vollmer et al., 2020) and C-ungrouped meteorite Tagish Lake (Herd et al., 2011). In Fig. 6 we compare our correlated Ryugu and Orgueil data to results from our laboratory acquired on polished CR (QUE 99177) and CM (Asuka 12236, Asuka 12169, Paris) meteorite sections using almost identical NanoSIMS methods to those used here (Nguyen et al., 2010; Nittler et al., 2021). Note that these CR and CM chondrites are the least altered specimens of their respective classes and all show much lower degrees of aqueous alteration in their mineralogies than do Ryugu or Orgueil. Nevertheless, the overall distributions of outlier D/H and $^{15}N/^{14}N$ ratios are very similar to each other, with only a small number of CR/CM grains showing higher D or $^{15}N$ enrichments than the most extreme Ryugu grains.

The Ryugu/Orgueil C-isotopic outliers are compared in Fig. 7 with data for CM, CR, and CO chondrites. Again, the Ryugu data span quite similar ranges to the carbonaceous chondrite data, which also show no correlation between H, C, and N isotopes. However, two differences are apparent in Fig 7A. First, the $^{13}C$-rich grains from CM chondrites extend to higher $\delta^{15}N$ values than the other chondrites and Ryugu. We attribute this difference to statistics – the CM dataset is much larger than any of the others (e.g., ~11,000 ROIs compared to <900 defined here for Ryugu). Second, there is a population (dashed ellipse) of $^{13}C$-depleted grains in CRs with $\delta^{15}N > 1000$ that is not seen in other chondrite classes. Moreover, this composition has also been reported (grey crosses) for carbonaceous grains from comet Wild-2 (De Gregorio et al., 2010) and a likely cometary stratospheric dust particle (Floss et al., 2004), though it does not represent a common composition in such samples. Very little is known about the nature of these grains, though Floss et al. (2004) proposed an interstellar origin.

### *4.2. Implications for the origin and evolution of organic matter in the Solar System*

The origin of the macromolecular organic matter in primitive Solar System materials is uncertain and highly debated, with arguments made for both interstellar and nebular origins (e.g., see review by Alexander et al., 2017). Although the meteoritic OM is chemically very complex and heterogeneous on multiple spatial scales, similarities in chemical functionality of bulk macromolecular material across chondrite groups, together with trends in bulk elemental and isotopic composition, are broadly consistent with all asteroids having sampled a common initial macromolecular precursor, with differences arising from varying degrees of parent-body modification (Alexander et al., 2017; Alexander et al., 2007). This idea is supported by the Tagish Lake meteorite (Herd et al., 2011), for which different clasts that experienced different degrees of aqueous alteration show correlated isotopic and elemental compositions in their bulk macromolecular material. In particular, aqueous alteration has appeared to lower both the bulk D/H and H/C ratios and the number of D-rich hotspots in the Tagish Lake macromolecular material (Fig. 4), as well as to modify the molecular structure, making it more aromatic (Alexander et al., 2014). Similar results were found in hydrothermal alteration experiments on IOM from the CM chondrite Murray (Yabuta et al., 2007), but recent hydrothermal experiments on CM chondrite organic matter at the relatively low temperatures inferred for CM and CI aqueous alteration did not find significant effects on D/H ratios, at least within the timescale of the experiments (Laurent et al., 2022), showing that we do not yet fully understand the effects of alteration on the H-isotopic composition of meteoritic organic matter.

The most extreme isotopic outliers (e.g., δD >5,000‰, $\delta^{15}$N> 1000; Figs. 6, 7) likely require an origin at extremely low temperatures, either in the presolar interstellar medium or in the far outer Solar System (Busemann et al., 2006; Remusat et al., 2009). The lack of correlation between H, C, and N isotopic compositions of these grains points to a strong diversity in fractionation pathways and local formation environments. However, the large majority of the organic particles in a given sample have less extreme and largely homogeneous compositions consistent with bulk values. The fact that this is observed to be the case even for samples that have seen much lower degrees of aqueous alteration than Ryugu or the Orgueil parent body (e.g., CR and CM data in Figs. 6, 7) indicates that if a common organic precursor originally formed from isotopically diverse interstellar grains (as may be represented by the hotspots and coldspots), these were largely homogenized before being accreted onto forming asteroidal bodies. Such a homogenization process could be caused, for example, by heating and or irradiation processes as individual grains were transported by turbulence throughout the protoplanetary disk, as envisioned for ice grains by Ciesla and Sandford (2012). Alternatively, the bulk of the organic grains could have formed in the nebula by distinct processes from the isotopic outliers (Remusat et al., 2006). In such a case, the outliers may represent a "salting" of the nebula by a late addition of interstellar or outer Solar System organic grains and have no genetic relationship to the bulk macromolecular material. However, coordinated isotopic-chemical-microstructural analyses of meteoritic organics have generally found no systematic differences between the isotopically outlier and average grains (Alexander et al., 2017; De Gregorio et al., 2013; Le Guillou et al., 2014). This is also reflected here in the correlated XANES-NanoSIMS results of Ryugu particle C0002 (Fig. 8). The six highlighted organic ROIs are more $^{15}$N-rich than typical Ryugu organics yet have fairly typical C-XANES spectra for Ryugu and CI chondrite macromolecular material. This suggests a genetic relationship between the outliers and the typical grains and supports the idea that, in bulk, the meteoritic organics reflect a largely, but not completely, homogenized sample of diverse interstellar grains, where the homogenization occurred both in the protoplanetary disk prior to accretion and during asteroidal hydrothermal processing.

Based on bulk Fe and Ni isotopic compositions, Hopp et al. (2022) have suggested that the Ryugu and CI parent bodies may have formed at a considerably further heliocentric distance than the parent bodies of other carbonaceous chondrites. If this was the case, one might expect an affinity with comets, which also are thought to have accreted much farther from the Sun. In this regard, it is interesting that no organic ROIs have been observed in the Ryugu or Orgueil samples with $^{13}$C depletions and very large $^{15}$N enrichments (ellipse in Fig. 7A) as this is a signature seen in some cometary samples (and possibly comet-related CR chondrites; Van Kooten et al., 2016). However, this difference may be due to the relatively small sample sizes of our datasets, relative to comparable datasets from other meteorites where large isotopic outliers are detected. In contrast, the bulk δD/H value of Ryugu and CI organics of ~500-1000‰ (Alexander et al., 2007; Remusat et al., 2022) is in the typical range of interplanetary dust particles (IDPs) thought to originate in comets (Alexander et al., 2017). However, such IDPs show no signs of aqueous alteration, whereas the D/H of Ryugu and CIs has likely been decreased by parent-body processing (Alexander et al., 2014), so the isotopic agreement may well be coincidental.

## 5. Conclusions

We analyzed by NanoSIMS the sub-micron distribution of H, C, and N isotopes in fragments of samples returned by the Hayabusa2 spacecraft from the primitive asteroid Ryugu as well as fragments of the Orgueil CI carbonaceous chondrite. We have found that:

1) The bulk H, C, and N isotopic compositions of the analyzed Ryugu fragments are in good agreement with the Orgueil samples analyzed under identical conditions. This provides additional support to there being a strong connection between Ryugu and the parent bodies of CI chondrites (Yokoyama et al., 2022).
2) The bulk isotopic compositions of the analyzed Ryugu fragments are also in good agreement with other Ryugu studies based on multiple analytical methods. The one notable exception is the study of Ito et al. (2022), which reported higher $^{15}N$ abundances on a ten-micrometer scale. This discrepancy may reflect heterogeneity within the returned samples.
3) We extracted H, C, and N isotopic compositions for C-rich ROIs within the NanoSIMS images for both Ryugu and Orgueil samples. Isotopic analysis is limited to grains larger than ~200 nm in diameter, a total of ~1300 ROIs each for Ryugu and Orgueil. We estimate that these ROIs represent about 25% of the total organic C in the samples; the remainder is in smaller grains and "diffuse" C associated with phyllosilicates. Again, the isotopic distributions of the Ryugu ROIs are essentially identical to those of the Orgueil samples.
4) Most of the C-rich ROIs show isotopic compositions within errors of each other and the bulk values. A small fraction of ROIs (≤9.6% for D/H, ≤6.8% for $^{15}N/^{14}N$, ≤2.7% for $^{13}C/^{12}C$; Table 2) are outliers, showing more extreme isotopic enrichments ("hotspots") or depletions ("coldspots"). The bulk abundances of the $^{15}N$-anomalous (0.15 – 0.44%) and $^{13}C$-anomalous grains (0.022 – 0.09 %) in the samples are comparable to those reported for hydrated Ryugu matrix by Nguyen et al. (2023). Although based on a very small area, the high abundance of $^{15}N$-rich grains observed in the FIB section from Ryugu C0002 (Fig. 8) of ~11% is consistent with the high abundance seen in a primitive clast from this particle (Nguyen et al., 2023).
5) The H, C and N isotopic compositions of outliers are not correlated with each other: while some C-rich grains are both D- and $^{15}N$-enriched, many are enriched or depleted in one or the other system. This most likely points to a diversity in isotopic fractionation pathways and thus diversity in the local formation environments for the individual outlier grains.
6) The observation that most individual sub-micrometer organic grains within a given meteorite have essentially the same isotopic compositions suggests either that they are a homogenized population of isotopically diverse precursors represented by the isotopic outliers or have a distinct origin from the outliers. The strong chemical similarity of isotopically typical and isotopically outlying grains, as reflected by XANES spectra, suggests a genetic connection and thus favors the former, homogenization scenario. However, the fact that even the least altered meteorites show the same pattern of a small population of outliers on top of a larger population of homogenized grains indicates that some or most of the homogenization occurred prior to accretion of the macromolecular organic grains into asteroidal parent bodies.

## Acknowledgements

LRN thanks NASA for supporting this work through the Hayabusa2 Participating Scientist Program (grant #NNX16AK72G). LR is grateful for funding from the European Research Council through the consolidator grant HYDROMA (grant agreement no. 819587). ZM acknowledges the financial support of Fundação para a Ciência e Tecnologia (FCT) (UIDB/00100/2020, UIDP/00100/2020, and LA/P/0056/2020). We thank Dr. Christian Vollmer and an anonymous referee for helpful comments which helped us to improve the paper.

**Supplementary Information** for *Microscale Hydrogen, Carbon, and Nitrogen Isotopic Diversity of Organic Matter in Asteroid Ryugu* by L. R. Nittler et al.

In addition to this file, two comma-separated-variable spreadsheets are provided:

NittlerSuppTable1.csv contains the isotopic data for individual ROIs reported in the main paper.

NittlerRyuguOrganicsSuppTable2.csv contains the C-XANES and NanoSIMS isotopic data for individual grains in a FIB section of Ryugu particle C0002, as shown in Fig.8 of the main paper.

## ROI definition

Figure S1 illustrates the ROI definition procedure we have used as applied to one of the Ryugu NanoSIMS images.

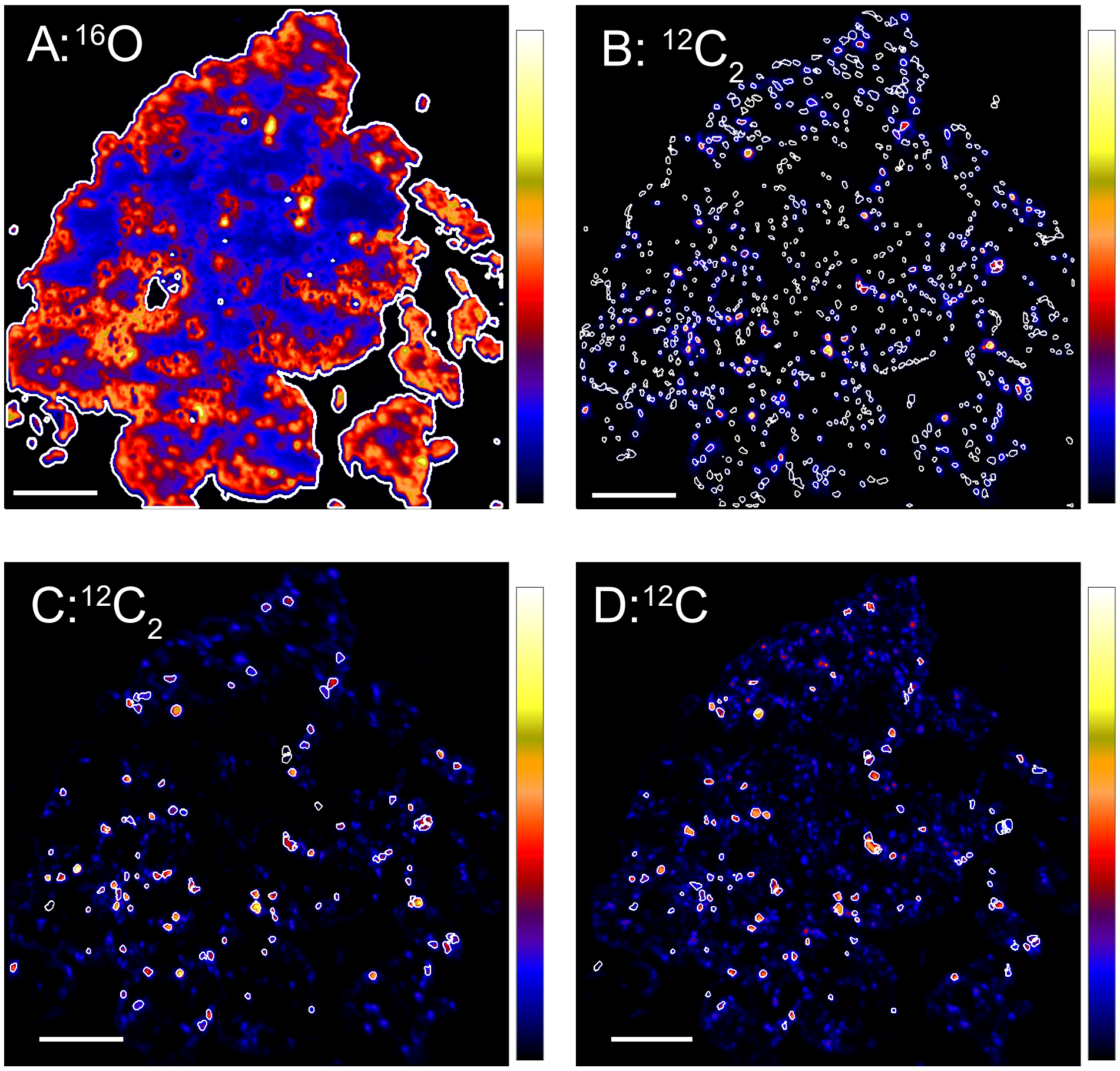

**Figure S1:** Example ROI definition for image HY2_A4-02a (Table 1). A) Oxygen-16 image with outline of "bulk" ROI. B) $^{12}C_2$ image with all C-rich ROIs defined. C) $^{12}C_2$ image with defined C-rich ROIs filtered to only include those with sufficient counts for isotopic measurements. D) $^{12}C$ image acquired during "H session" with aligned ROI definition. Scale bars are 5 microns.

## C-isotope corrections

After extracting isotopic data for the C-rich ROIs in all of the images, we discovered there is a relationship between the $^{12}C_2^-$ count rate and measured $^{13}C/^{12}C$ ratio for the ROIs from both the Ryugu and Orgueil NanoSIMS images (Fig. S2A). This most likely reflects "quasi-simultaneous arrivals," in which more than one secondary ion are emitted when a single primary ion hits the sample surface and thus are registered as a single pulse in the ion counting system (Ogliore et al., 2021; Slodzian et al., 2004). This leads to undercounting of abundant species with high ionization efficiency, like $C_2$ from organic matter. We empirically corrected for this effect in the data by following the formalism of Slodzian et al. and calculating a corrected $^{12}C_2^-$ count rate for each ROI such that the slope of the relationship becomes flat (Fig. S2B). The y-intercept of the trend at -45‰ gives the average measured $\delta^{13}C$ value for the data. We assume that the true composition of both Ryugu and Orgueil organics is the same as insoluble organic matter from CI chondrites (-17 permil; Alexander et al., 2007); this assumption is supported by Ryugu data reported by Nakamura et al. (2022). This gives an instrumental fractionation factor of -28 ‰, which was used to correct the ROI data.

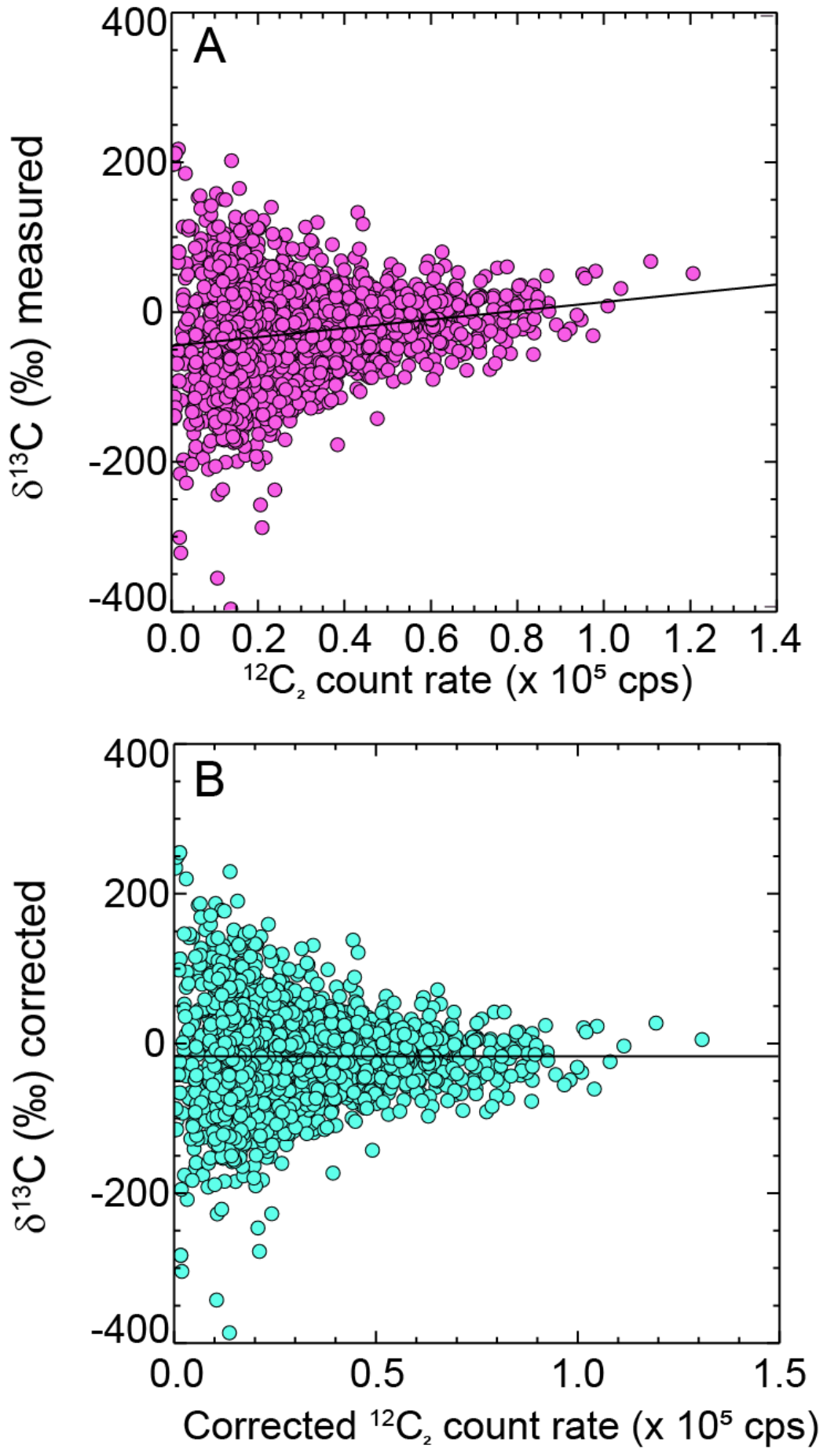

**Figure S2.** A) Uncorrected and B) corrected C-isotope data for Ryugu and Orgueil ROIs (see text).

**Data Filtering**

Once ROIs had been defined and isotopic ratios extracted for them, we examined the resulting data to determine which were statistically significant enough to be included in the plots and discussion. The uncertainty on ROI isotopic ratios is governed by Poisson statistics and thus increases rapidly as the number of counted ions goes down. In general, if one plots the measured ratio versus number of counts one expects to see the distribution of points broaden smoothly as the counts decrease. However, below a certain count rate, the number of ions of the minor isotope can reach very small numbers (e.g. single digits) where the asymmetry of the Poisson distribution dominates and reliable data cannot be easily derived. To illustrate this, Fig S3 shows the measured $\delta^{15}N$ values for our Ryugu ROIs plotted against the total number of denominator ($CN^-$) counts:

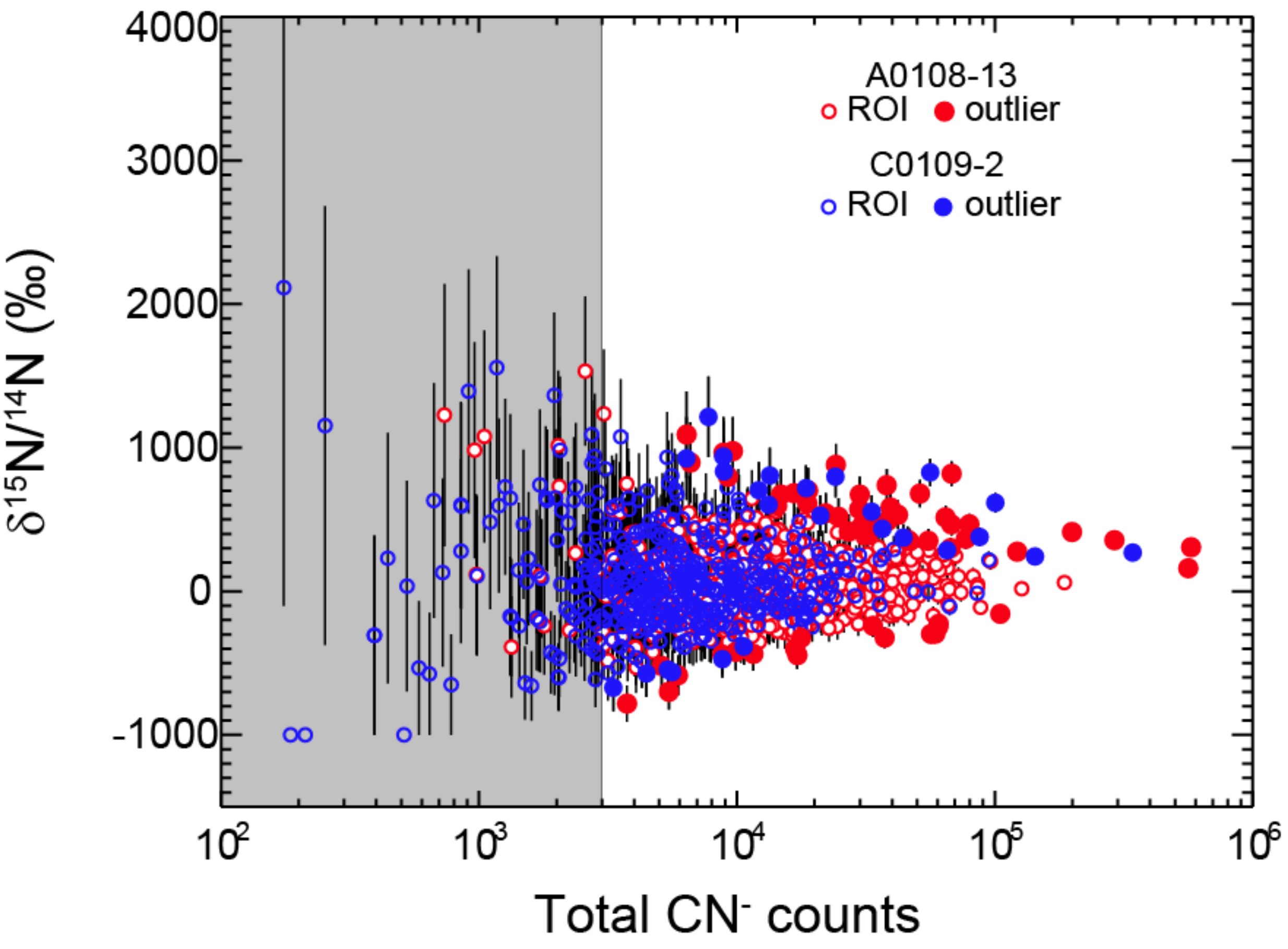

**Fig. S3.** See text

At relatively CN high counts (>several $10^4$), the main distribution (open circles) of $\underline{\delta}^{15}N$ values is relatively narrow, but widens smoothly and symmetrically as CN counts decrease. Filled symbols indicate ROIS that are >3σ from the average of the distribution and thus are identified as “outliers.” Note that whether an outlier can be identified depends on the count rate and thus depends both on the grain size and the N/C ratio (for N isotopes). Below a CN count of ~3 x $10^3$ (gray area), the distribution greatly widens and no more outliers can be identified. We thus take this value as a threshold and only consider ROIs with CN counts higher than this for our discussions. Similarly, we derive a threshold of $H^-$ counts > 6 x$10^3$ in a given ROI in order to consider a measured δD value significant.